\documentclass[%
reprint,
 amsmath,amssymb,
aps,
]{revtex4-1}

\usepackage[dvipdfmx]{graphicx} 
\usepackage{bmpsize}
\usepackage{dcolumn}
\usepackage{bm}
\usepackage{natbib}
\usepackage{longtable}
\usepackage{adjustbox}

\makeatletter
\newcommand*{\rom}[1]{\expandafter\@slowromancap\romannumeral #1@}
\makeatother
\begin{document}


\title{Laser-induced dissociative ionization of H$_{2}$ from the near-IR to the mid-IR regime}

\author{Qingli Jing}
\author{Lars Bojer Madsen}%
\affiliation{Department of Physics and Astronomy, Aarhus University, 8000 Aarhus C, Denmark
}%




\date{\today}

\begin{abstract}
We apply the Monte Carlo Wave Packet (MCWP) approach to investigate the kinetic energy release (KER) spectra of the protons following double ionization in H$_{2}$ when interacting with laser pulses with central wavelengths ranging from the near-IR (800 nm) to the mid-IR (6400 nm) regions and with durations of 3-21 laser cycles. We uncover the physical origins of the peaks in the nuclear KER spectra and ascribe them to mechanisms such as ionization following a resonant dipole transition, charge-resonance-enhanced ionization (CREI) and ionization in the dissociative limit of large internuclear distances. For relatively large pulse durations, i.e., for 15 or more laser cycles at 3200 nm and 10 or more at 6400 nm, it is possible for the nuclear wave packet in H$_{2}^{+}$ to reach very large separations. Ionization of this part of the wave packet results in peaks in the KER spectra with very low energies. These peaks give direct information about the dissociative energy in the $2p\sigma_{u}$ potential energy curve of H$_{2}^{+}$ at the one- and three-photon resonances between the $2p\sigma_{u}$ and $1s\sigma_{g}$ curves in H$_{2}^{+}$. With the MCWP approach, we perform a trajectory analysis of the contributions to the KER peaks and identify the dominant ionization pathways. Finally, we consider a pump-probe scheme by applying two delayed pulses to track the nuclear dynamics in a time-resolved setting. Low-energy peaks appear for large delays and these are used to obtain the $2p\sigma_{u}$ dissociative energy values at the one-photon resonance between the $2p\sigma_{u}$ and $1s\sigma_{g}$ curves in H$_{2}^{+}$ for different wavelengths.
\begin{description}
\item[PACS numbers] 33.20.Xx, 33.40.+f, 33.80.Eh,
\end{description}
\end{abstract}
\pacs{Valid PACS appear here}
\maketitle

\section{\label{sec:level1} Introduction }
The significant advances in femtosecond laser technology have opened the possibility to control chemical reactions with laser light \cite{zhu1995coherent,zewail2000femtochemistry,hertel2006ultrafast}. When exposing molecules to intense laser pulses, strong interaction between molecules and the external electromagnetic field may give rise to electronic excitation or even result in ionization. Nuclear dynamics is hence often induced because of the interplay between the nuclei and electrons. A good example is the process of dissociative ionization \cite{rapp1965cross,codling1993dissociative}, a process in which dissociation is induced as a result of the removal of one or more electrons. Laser-induced dissociative ionization has received considerable interests in the past two decades \cite{chelkowski1998electron,posthumus2004dynamics,yue2013dissociation}. Thanks to the fast developments in imaging techniques such as COLTRIMS \citep{PhysRevLett.93.163401} and VMI spectrometers \cite{struder2010large}, it is now possible to measure the kinetic energy of the ionic fragments or the liberated electrons or both. By analysing the kinetic energy release (KER) spectra of the ionic fragments, information about the ionization process can be obtained \cite{thumm2008time}. For example, both the process of charge-resonance-enhanced-ionization (CREI) \cite{bandrauk1999charge,zuo1995charge} and resonance-enhanced multiphoton ionization \cite{yang1991high,apalategui2002multiphoton} manifest themselves through the occurrence of characteristic peaks in the nuclear KER spectra \cite{chelkowski2007dynamic}. Different from the mechanism of resonance-enhanced multiphoton ionization, for one-electron system like H$_2^+$, CREI results from lowering the internal barrier between the double wells  at certain critical internuclear separations in the presence of a strong field, along with the creation of a pair of field-dressed charge-resonant states.\par

In parallel to the advances in experimental techniques, many theoretical methods \cite{PhysRevA.71.012712,PhysRevA.73.033410,Calegari336} have been developed and proven to work very well in describing a range of phenomena associated with the interaction of molecules and light. Among these methods, the Monte Carlo wave packet (MCWP) approach \cite{dalibard1992wave,leth2009monte} performs well when describing the laser-induced dissipative dynamics. Especially in previous works on H$_{2}$ \cite{leth2010monte,leth2010dissociative}, and O$_{2}$ \cite{leth2011dissociative}, where the nuclear KER spectra were of main concern, the simulations agreed very well with the measurements. This approach, like many others, is devoted to solve the time-dependent Schr\"{o}dinger equation, but the problem is greatly simplified as the electronic degrees of freedom are treated in an effective description. At the same time, the Markov approximation \cite{breuer2016colloquium} is applied, which means that the flow of electrons from the molecular systems to the surroundings is uni-directional and irreversible. The loss of electrons is encoded in jump operators, which are responsible for jumps among different Hilbert spaces and constitute a non-Hermitian part of the total Hamiltonian. The non-Hermitian Hamiltonian implies that the norm of the system is decreasing over time. The drop of the norm over time is controlled by the instantaneous ionization rates \cite{vafaee2006detailed, yudin2001nonadiabatic} of the system. If the ionization rates depend on the nuclear configuration, wave packet motion is induced in the neutral molecule. Inducing nuclear dynamics in this way is called Lochfra{\ss} \cite{PhysRevLett.97.103003,PhysRevLett.97.103004,Asaenz2016arxiv} and this phenomena is an integrated part of the MCWP simulations. In the present MCWP approach to dissociative ionization, the jumps between different charge states and associated Hilbert spaces are described by ionization rates. While the evolution of the system within a given charge state is coherent, the rate treatment of the ionization step means that coherence between the remaining cation and the ionized electron is lost. Also we have no knowledge of the energy distribution of the outgoing electron. \par 

Recently there has been a shift in strong-field physics towards the use of mid-IR driving wavelengths \cite{wolter2015strong}, e.g., to image molecular structures by light-induced electron diffraction (LIED) \cite{blaga2012imaging,meckel2008laser} and to greatly increase the photon energies of high-harmonic generation (HHG) \cite{popmintchev2012bright}. The interaction between the mid-IR laser pulses and molecules can be treated in the quasi-static regime. 
To our knowledge, the behavior of the nuclear KER spectra with increasing wavelength from the near-IR region to the mid-IR region has not been studied systematically. Such a systematic investigation of the trends in the spectra and the underlying physics when going to longer wavelengths is the topic of the present work. The large wavelengths of the intense IR laser pulses make the ionization dynamics of molecules enter the regime of tunneling ionization \cite{tong2002theory,urbain2004intense,PhysRevA.84.053423,PhysRevA.93.053426}. Studying the peaks in the KER spectra at mid-IR wavelengths can help to obtain detailed knowledge of the nuclear motion, as will be discussed below. As a characteristic trend in the spectra, we see a shift towards lower KER values with increasing wavelength. \par

The paper is organized as follows. In Sec.~\ref{sec:level2}, we review how to apply the MCWP method to simulate dissociative ionization in H$_{2}$. In Sec.~\ref{sec:level3}, we discuss the KER spectra obtained for H$_{2}$ exposed to laser pulses with different wavelengths and pulse durations. In Sec.~\ref{sec:level4}, the nuclear KER spectra of H$_{2}$ interacting with pump-probe pulses with delays are analyzed. Sec.~\ref{sec:level5} concludes. Atomic units ($\hbar =e=m_{e}=a_{0}=1$) are used throughout unless stated otherwise.
 
\section{\label{sec:level2} MCWP approach for dissociative ionization of H$_{2}$}
\begin{figure}[!ht]
  \includegraphics[width=0.40\textwidth,natwidth=610,natheight=500]{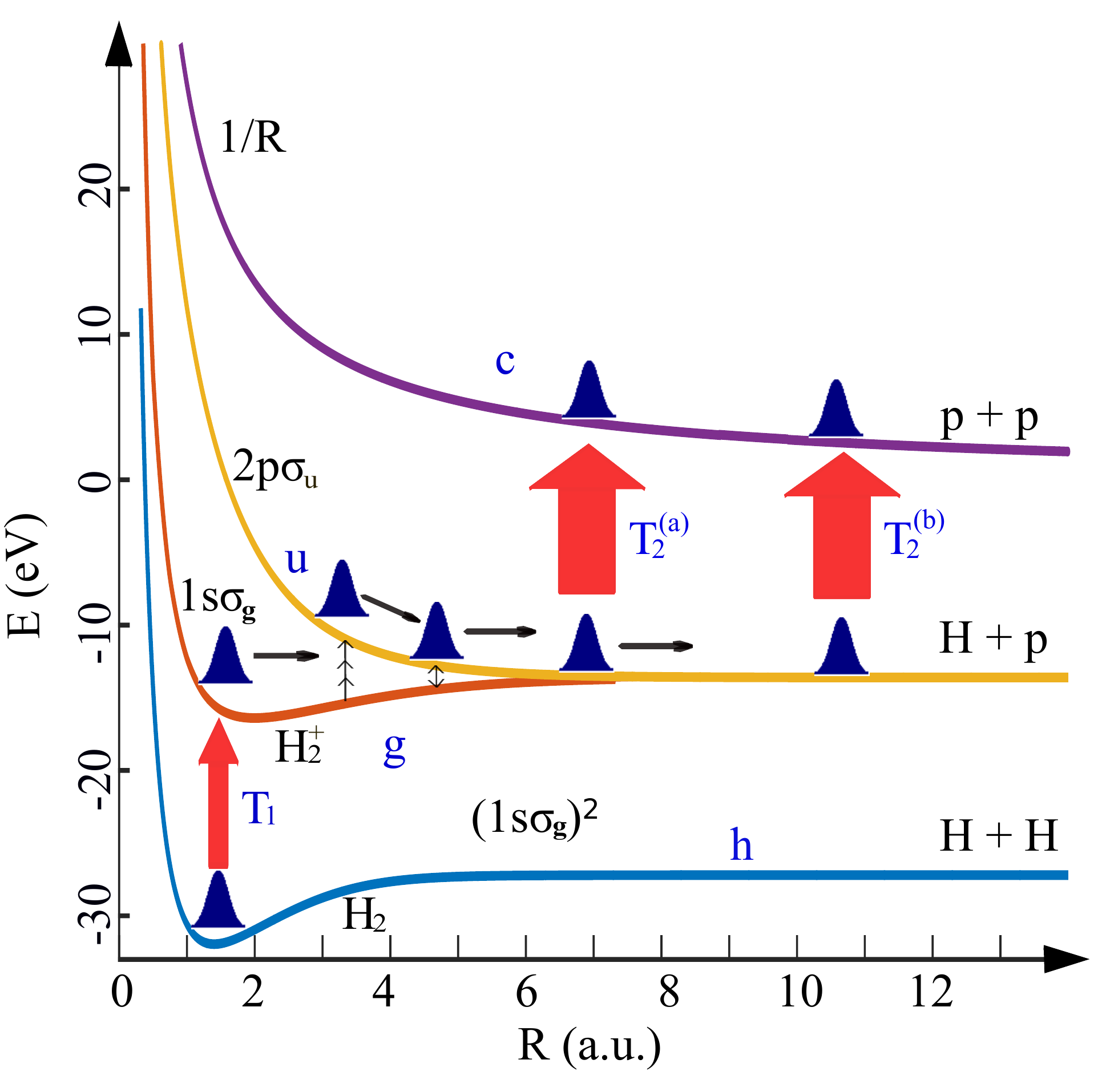}
  \centering
  \caption{ Sketch of four field-free Born-Oppenheimer electronic potential energy curves \cite{lindstrom2005nist}. These curves are the $(1s\sigma_{g})^2$ ground-state curve in H$_{2}$ labeled by $h$, the $1s\sigma_{g}$ and $2p\sigma_{u}$ curves in H$_{2}^{+}$ labeled by $g$ and $u$, respectively, and the $1/R$ Coulombic curve labeled by $c$. A particular realization of a quantum trajectory with the MCWP approach is also shown: At some instant $T_{1}$, the first ionization occurs and the neutral wave packet is promoted to the singly-ionized system where coherent evolution takes place. At some later instant $T_{2}^{j}$ ($j=a$,$b$), the second ionization occurs and then the nuclei undergo Coulomb repulsion. The final energy of the protons depends on the instants $T_{1}$, $T_{2}^{j}$. }  
  \label{fig:potential_curves}
\end{figure}
The MCWP approach for dissociative ionization was described in detail elsewhere \cite{leth2009monte, leth2010monte,leth2010dissociative,leth2011dissociative,henriette}, so the description here is brief. The dissipative dynamics of a diatomic molecule interacting with laser light can be described by the time-dependent Schr{\"o}dinger equation (TDSE)
\begin{equation}
i\partial_{t}|\Psi(t)\rangle=H(t)|\Psi(t)\rangle=(H_{s}-\frac{i}{2}\sum_{m}C_{m}^{+}C_{m})|\Psi(t)\rangle,
\label{equation:TDSE}
\end{equation}
with the Hermitian Hamiltonian 
\begin{equation}
H_{s}=T_{N}+V_{N}+V_{eN}+T_{e}+V_{ee}+V_{I}(t),
\end{equation}
where $T_{N}$ is the kinetic energy operator for the nuclei, $T_{e}$ the kinetic energy for the electrons, $V_{N}$ the nuclear repulsion, $V_{eN}$ the nuclei-electron interaction, $V_{ee}$ the electron-electron interaction and $V_{I}(t)$ the laser-matter interaction. 
Within the Born-Oppenheimer approximation, the total state $|\Psi(t)\rangle$ can be expanded as 
\begin{equation}
|\Psi(t)\rangle=\sum_{m}\int d\vec{R}X_{m}(\vec{R},t)|\phi^{el}_{R,m}\rangle\otimes |\vec{R}\rangle,
\label{equation:total states}
\end{equation}
in which $X_{m}(\vec{R},t)$ is the nuclear wave packet and $|\vec{R}\rangle$ refers to the position eigenkets of the nuclear coordinate. 
The electronic basis states $|\phi^{el}_{R,m}\rangle$ can be obtained by solving the time-independent Schr{\"o}dinger equation with parametric dependence on the internuclear separation $\vec{R}$:
\begin{equation}
(T_{e}+V_{ee}+V_{eN}+V_{N})|\phi^{el}_{R,i}\rangle=E_{el,i}({R})|\phi^{el}_{R,i}\rangle,
\label{equation:electronic basis}
\end{equation}
where $E_{el,i}({R})$ is the corresponding electronic potential energy curve. The four electronic potential energy curves included in our MCWP calculations are shown in Fig. \ref{fig:potential_curves}. These are the $(1s\sigma_{g})^2$ curve in H$_{2}$, the $1s\sigma_{g}$ and $2p\sigma_{u}$ curves in H$_{2}^{+}$, and the doubly-ionized Coulombic curve in H$_{2}^{++}$ - for short denoted $h$, $g$, $u$, and $c$, respectively. These curves were previously shown to be sufficient to capture the dynamics responsible for the KER spectra \cite{leth2010monte,leth2010dissociative}.
The jump operators $C_{m}$ in Eq. (\ref{equation:TDSE}) constitute a non-Hermitian term in the total Hamiltonian. This term is responsible for the jumps among the different Hilbert spaces of H$_{2}$, H$_{2}^{+}$ and H$_{2}^{++}$,
\begin{equation}
C_{m}=\int d\vec{R}\sqrt{\Gamma_{m}(\vec{R},t)}|\phi^{el}_{R,n}\rangle\langle\phi^{el}_{R,m}|\otimes |\vec{R}\rangle\langle\vec{R}|,
\label{equation:jump operators}
\end{equation}
where $\Gamma_{m}(\vec{R},t)$ is the ionization rate from the $m$ ($h$, $g$ or $u$) state to the $n$ ($g$, $u$ or $c$) state and depends on the instantaneous value of the field strength. In this work, the ionization rate responsible for the ionization of H$_2$ is calculated using the weak-field asymptotic theory of Ref. \cite{PhysRevA.84.053423}. The ionization rates for the ionization of the cation are obtained by interpolating the results published in Ref. \cite{plummer1996field}. \par
By substituting Eqs. (\ref{equation:total states}), (\ref{equation:electronic basis}) and (\ref{equation:jump operators}) into Eq. (\ref{equation:TDSE}) and 
using the ansatz $X_{m}(\vec{R},t)=\frac{1}{R}K_{m}(R,t)W_{m}(\theta,\phi,t)$, the time evolution of the radial wave function $K_{m}(R,t)$ can be obtained. As the molecule is assumed rotationally frozen during the interaction with the femtosecond laser pulse, the angular parts of the nuclear wave function $W_{m}(\theta,\phi,t)$ can be expressed as $W_{m}(\theta,\phi,t)=\frac{1}{\sqrt{4\pi}}\delta(\theta-\theta_{0})\delta(\phi-\phi_{0})$ with $\theta_{0}$ and $\phi_{0}$ specifying the internuclear orientation. In the calculations below, we consider the case where the molecule is aligned with the linear polarization of the external field, i.e.,  $\theta_0 = 0^\circ $ and $\phi_0=0^\circ$, where the latter can be arbitrary because of symmetry.
\par

The simulation procedure of the MCWP technique is outlined in the following (see also Refs. \cite{leth2010monte,leth2010dissociative}). The nuclear wave function first evolves in the neutral system with its norm square decreasing because of the non-Hermitian term in the Hamiltonian. At each time step, the drop in the norm square is compared with a random number (between 0 and 1), which determines whether the jump from one Hilbert space to another takes place or not. If the drop in the norm square is larger than the random number, ionization occurs and thus the wave function jumps, is renormalized and then evolves in the new Hilbert space, otherwise, ionization can not occur and the wave function needs to be renormalized in the neutral system. The comparison between the drop in the norm square and a random number continues until the first jump occurs. Similarly, whether the second jump takes place or not is controlled by comparing the drop in the norm square of the nuclear wave function in H$_{2}^{+}$ with a new random number (between 0 and 1). The difference compared to the first jump is that there are two pathways for the second jump to the doubly-ionized system, from the $g$ state curve and from the $u$ state curve, which means that another random number (between 0 and 1) should be introduced to decide which pathway to choose. As a result, a nuclear wave function after emitting two electrons is obtained stochastically, with the first jump occurring at some instant of time $T_{1}$ and the second jump at some instant of later time $T_{2}$ [Fig. \ref{fig:potential_curves}]. The nuclear wave packet after two jumps $K_{c}(R,T_{2})$ holds all the information of the nuclei. For example, the nuclear kinetic energies can be obtained through projecting the wave packet on Coulomb waves $K_{E}(R)$, 
\begin{equation}
\text{KER}_{m}(T_{1},T_{2})=|\int K_{E}(R)K_{c}(R,T_{2})dR|^{2},
\label{equation:ker}
\end{equation}
with $m=g$, $u$. Equation (\ref{equation:ker}) only gives the KER signal for one stochastic event occurring at specific $T_{1}$ and $T_{2}$. We refer to such a specific realization of jump times and dynamics as a quantum trajectory. The total nuclear KER spectra of the dissociative ionization process can be obtained by averaging over all the stochastic events. In the present case of dissociative double ionization, the computational effort can, however, be dramatically reduced through eliminating all the random numbers by applying a completely deterministic approach. To this end, the two jumps which are dressed with probabilities are assumed to take place at every time step. Actually, the probability for each jump is proportional to the drop in the norm square at that jump time. This is reasonable as the larger the drop in the norm square, the more likely it is for a jump to occur. Summing over all the weighted events, the nuclear KER spectra can be obtained from the formula, 
\begin{equation}
\text{KER}_{\text{tot}}=\sum_{T_{1},T_{2}}P_{1}P_{12}\sum_{m=g,u}P(m|\{T_{1},T_{2}\})\text{KER}_{m}(T_{1},T_{2}).
\label{equation:kertot}
\end{equation} 
In the above equation, $P_{1}=-{d}(N_{h}(t))/{dt}$ is the probability density for the first jump with $N_{h}(t)$ representing the norm square of the nuclear wave function of the $h$ state; $P_{12}=-{d}(N_{g}(t)+N_{u}(t))/{dt}$ is the conditional probability density of the second jump for a given first jump with $N_{g}(t)$ and $N_{u}(t)$ representing the norm square of the nuclear wave function of the $g$ state and the $u$ state, respectively; and $P(m|\{T_{1},T_{2}\})={\langle \Psi|C_{m}^{+}C_{m}|\Psi\rangle}/{\langle \Psi|C_{g}^{+}C_{g}+C_{u}^{+}C_{u}|\Psi\rangle}$ is the conditional probability density from the $m$ ($g$ or $u$) state for given first and second jump times (see also Ref. \cite{henriette}).\par

As shown in Fig. \ref{fig:potential_curves}, the initial nuclear wave function is assumed to be ground state of the $(1s\sigma_{g})^{2}$ potential energy curve of H$_{2}$. The applied laser pulse with polarization axis parallel to the molecular axis first kicks out the first electron and then induces nuclear motion in the two potential energy curves of the H${_{2}^{+}}$ system. Meanwhile, there is electronic dipole coupling between the $1s\sigma_{g}$ and $2p\sigma_{u}$ states. Finally, the nuclei experience Coulomb repulsion after emitting the second electron. Thus the nuclei pick up energy from two terms, one is the dissociative kinetic energy in the singly-ionized system and the other is the Coulomb repulsion energy in the doubly-ionized system. The time-dependent Hamiltonian, including the non-Hermitian part representing the interaction between states in different Hilbert spaces as well as the electronic dipole coupling between states within a given Hilbert space, originates from the external laser field. As a result and as shown in the next section, the kinetic energy is closely related to the parameters of the applied laser pulse, such as peak intensity, wavelength, and pulse duration. \par

Numerically, within each Hilbert space, we solve the TDSE by applying the split-operator method \cite{FEIT1982412} on the short-time propagator, i.e.,
\begin{eqnarray}
U(t+\Delta t,t)&=&\text{exp}(-iT_{N}\frac{\Delta t}{2})\text{exp}(-iV({R},t+\frac{\Delta t}{2})\Delta t) \nonumber \\
&\times&\text{exp}(-iT_{N}\frac{\Delta t}{2}).
\label{equation:short-time operator}
\end{eqnarray}
Obviously, it is the term $V(R,t)$ that determines how the wave packet evolves in each Hilbert space. For example, in H$_{2}^{+}$, the 2-by-2 matrix representation of $V(R,t)$ is $V_{ij}(R,t)=E_{el,i}(R)\delta_{ij}-i\Gamma_{i}({R},t)\delta_{ij}+\beta D_{ij}(R)F(t)$ ($i,j=g,u$), where $\beta=1+1/(2m_{\text{p}}+1)$ with $m_{\text{p}}$ the proton mass, $F(t)$ is the electric field, and $D_{ij}(R)$ is the electronic dipole moment function between $i$ and $j$ states along the direction, $\hat{\epsilon}$, of the linear polarization, i.e., $D_{ij}(R)=-\langle \phi^{el}_{R,i}|\vec{r}\cdot\hat{\epsilon}| \phi^{el}_{R,j}\rangle$. We use the explicit expression for $D_{gu}( R)=\frac{R}{2\sqrt{1-((1+R+R^2/3)e^{-R})^2}}-\frac{1}{(2+1.4R)}$ given in Ref. \cite{PhysRevA.63.013408}. The size of our simulation box is 40.96. The time step $\Delta t$ is 0.1 and the spatial step $\Delta R$ is 0.02. The operators $\text{exp}(-iT_{N}\frac{\Delta t}{2})$ and $\text{exp}(-iV({R},t+\frac{\Delta t}{2})\Delta t)$ are diagonal in the momentum and position representations, respectively, thus a fast-Fourier-transform algorithm is applied in the implementation.
 
\section{\label{sec:level3} WAVELENGTH AND PULSE DURATION DEPENDENCE OF KER spectra for H$_{2}$ interacting with a single laser pulse}
\begin{figure}[!ht]
  \includegraphics[width=0.4\textwidth,natwidth=450,natheight=530]{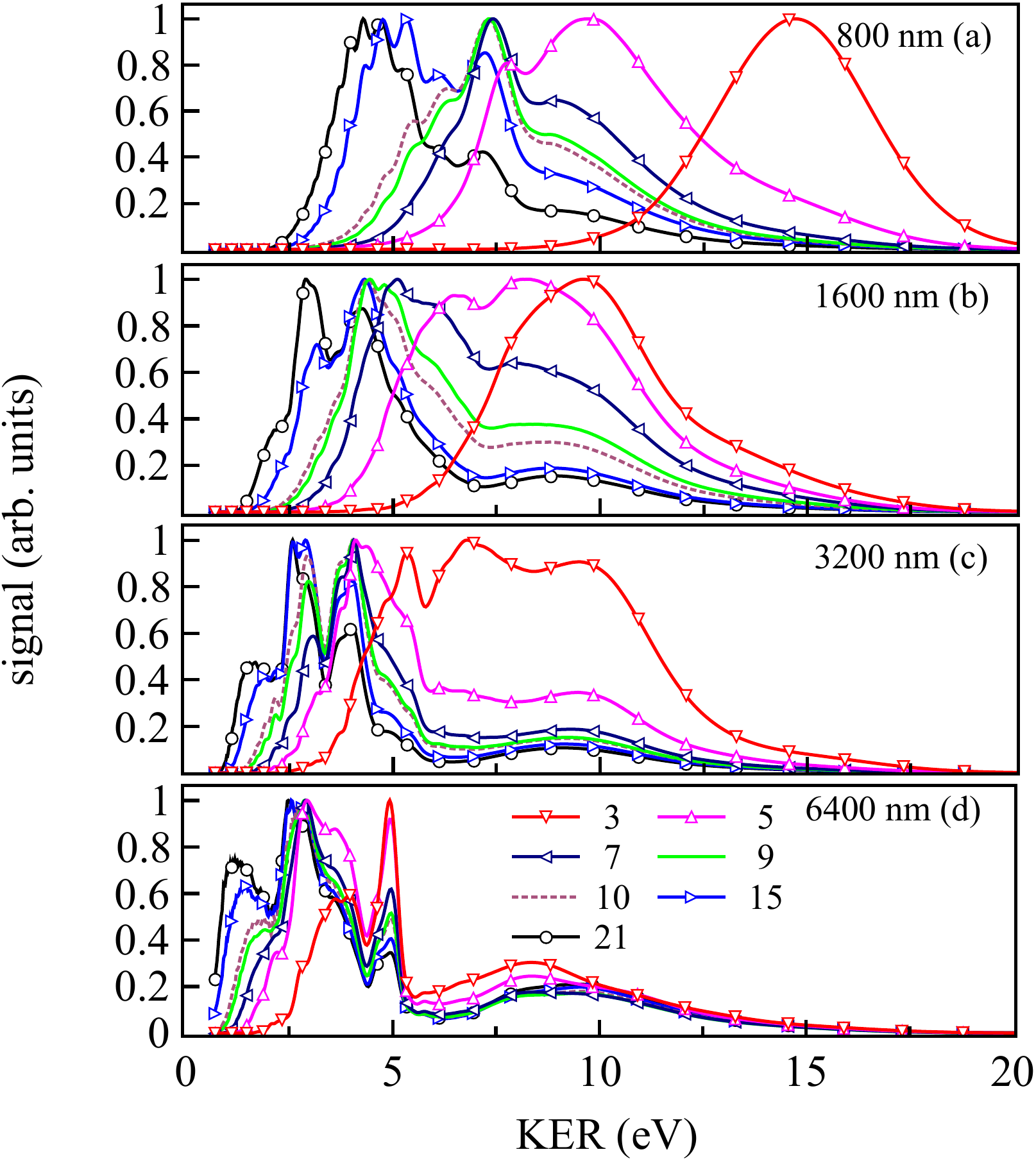}
  \includegraphics[width=0.4\textwidth,natwidth=450,natheight=170]{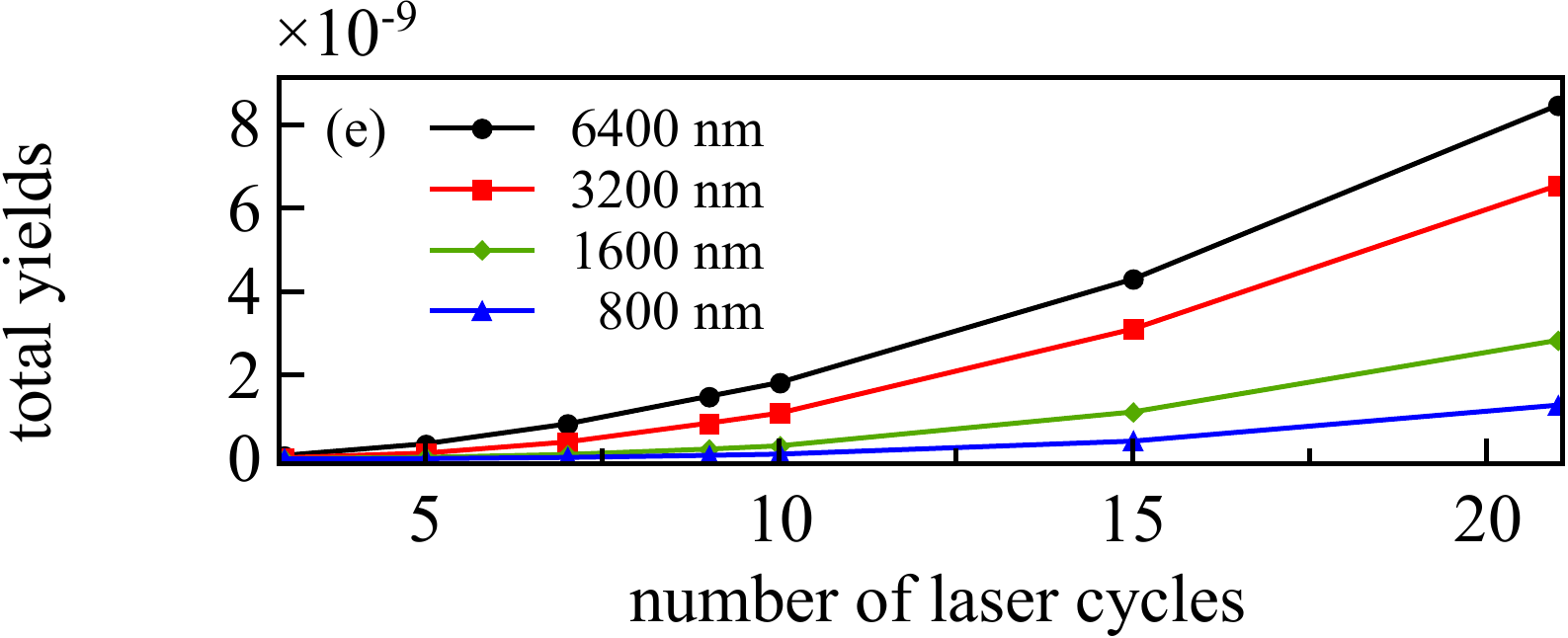}
  \centering
  \caption{ (a)-(d) Renormalized nuclear KER spectra after double ionization of H$_{2}$ interacting with laser pulses with different wavelengths ranging from 800 nm to 6400 nm and with different FWHM pulse durations ranging from 3 to 21 laser cycles. All the pulses have a peak intensity of $6\times 10^{13}$ $\text{W}/ \text{cm}^{2}$. (e) Total yields of protons in units of ionization probability per molecule as a function of the number of laser cycles for several wavelengths.} 
  \label{fig:kertot_wavelength_cycles}
\end{figure}

In this section, we simulate the process of dissociative double ionization of ground-state H$_{2}$ interacting with single laser pulses. The calculated proton KER spectra for different wavelengths (ranging from 800 nm to 6400 nm) and pulse durations (ranging from 3 to 21 laser cycles) are shown in Fig. \ref{fig:kertot_wavelength_cycles}. In our calculation, the laser pulses are of Gaussian shape, and the pulse durations are expressed in terms of the number of laser cycles within the FWHM of the intensity profile of the applied laser pulses. The laser intensity of $6\times 10^{13}$ $\text{W}/ \text{cm}^{2}$ is chosen to be similar to the one in the experiment of Ref. \cite{PhysRevLett.98.073003}. If there are specific internuclear positions from which a significant part of the second ionization takes place, distinctive peaks occur in the nuclear KER spectra. These peaks mostly come from events in H$_{2}^{+}$ which occur at internuclear separations with a resonant one- or three-photon dipole coupling between the $1s\sigma_{g}$ and $2p\sigma_{u}$ curves, or at the separations of CREI \cite{bandrauk1999charge,zuo1995charge}, or in the dissociative limit beyond the CREI distances. When we look at Fig. \ref{fig:kertot_wavelength_cycles} from the top to the bottom, we see in general that the KER spectra move towards lower kinetic energies with increasing duration of the laser pulses. This behavior is because the larger the duration of the pulse, the more likely it is for the nuclei to dissociate to larger separations with correspondingly smaller repulsive Coulomb energy. Moreover, when interacting with laser pulses with the same duration, for example 10 laser cycles at 800 nm and 5 laser cycles at 1600 nm, even though the individual peaks are shifted because of the different positions for the resonant electronic dipole coupling in H$_{2}^{+}$, the overall widths of the KER spectra windows look quite similar. We will now identify the most prominent peaks in the spectra in Fig. \ref{fig:kertot_wavelength_cycles}. The special KER peak around 9 eV visible in Figs. \ref{fig:kertot_wavelength_cycles}(a-d), is a signature of the outer turning point of the nuclear wave packet evolving along the $1s\sigma_{g}$ BO potential energy curve in H$_2^{+}$. The peak is nearly independent of the laser pulse and is in nice agreement with the classical estimate of the Coulomb repulsion energy of $1/R_{tp} \sim 1/3$ (9.1 eV), with $R_{tp}\sim 3$ the outer turning point of the wave packet along the $1s\sigma_{g}$ potential energy curve, when the initial nuclear wave packet is assumed to be that of the ground state of the $(1s\sigma_{g})^2$ curve with an equilibrium internuclear separation of 1.4. \par

For a fixed wavelength, several peaks in the KER spectra are more or less at similar positions for different numbers of laser cycles within the FWHM intensity duration. This behavior is a signature of the fact that the positions of the resonant electronic one- and three-photon dipole coupling are the same for these laser pulses. Moreover, there are clear signatures of CREI \cite{zuo1995charge} in the spectra. CREI takes place at internuclear separations around $R=7$ and $R=11$ \cite{plummer1996field} at this laser peak intensity, and gives rise to Coulomb repulsive energies of 3.9 eV and 2.5 eV. When the wavelength is larger, the distances between positions for the one- or three-photon resonance and positions for CREI are smaller, and thus the dissociative kinetic energy picked up along the $2p\sigma_{u}$ curve before the CREI positions is becoming smaller. The peaks around 3 eV (smaller than the Coulomb repulsion at $R=7$) in Figs. \ref{fig:kertot_wavelength_cycles}(b-d) is a signature of CREI around $R=11$, while in Fig. \ref{fig:kertot_wavelength_cycles}(a) the pulse durations are too short to allow a substantial portion of the nuclear wave packet to reach $R=11$ during the pulse.\par 
 \begin{figure}[!ht]
  \includegraphics[width=0.45\textwidth]{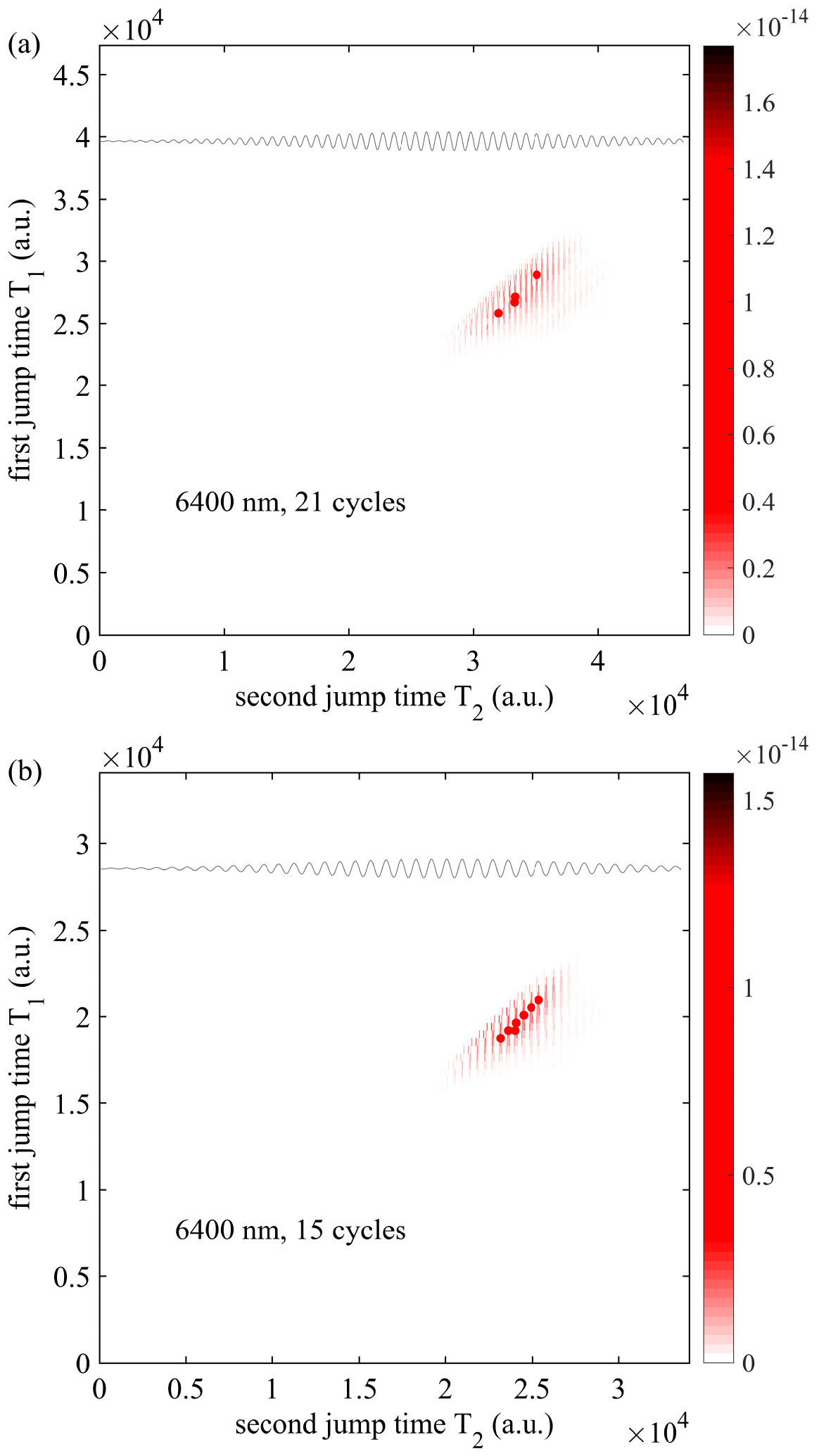}
  \centering
  \caption{(a) Contributions from different trajectories to the low-energy peak around 1.2 eV in Fig. \ref{fig:kertot_wavelength_cycles}(d) when interacting with a pulse with 21 laser cycles at 6400 nm. (b) Contributions from different trajectories to the low-energy peak around 1.5 eV in Fig. \ref{fig:kertot_wavelength_cycles}(d) when interacting with a pulse with 15 laser cycles at 6400 nm. The red dots in (a) and (b) mark the trajectories which give the largest contributions. The laser pulses are indicted by the grey lines and the parameters are as in Fig. \ref{fig:kertot_wavelength_cycles}(d).} 
  \label{fig:kertot_1ev_6400nm}
\end{figure}

When interacting with laser pulses with larger wavelengths, see Figs. \ref{fig:kertot_wavelength_cycles}(c) and \ref{fig:kertot_wavelength_cycles}(d), some pronounced new energy peaks appear around 1.5 eV when the number of laser cycles is larger than 15 (10) for 3200 nm (6400 nm). These peaks occur because the nuclear wave packet can move further during the longer pulse and be ionized at separations beyond the positions of CREI. These low-energy peaks are of the same origin as the $<2$ eV peak monitored in a previous pump-probe experiment \cite{niikura2006controlling}. The low-energy peaks for 15 and 21 laser cycles at 6400 nm are at about 1.5 eV and 1.2 eV, respectively. We benefit from the stochastic sampling within the MCWP approach and carry out a trajectory analysis of these two peaks. The contributions from different trajectories are shown in Fig. \ref{fig:kertot_1ev_6400nm}. The ionization rate from the $2p\sigma_{u}$ state to the doubly-ionized molecule is at least one order of magnitude larger than that from the $1s\sigma_{g}$ state for $R\geq 3$ \cite{plummer1996field}. Moreover, only a very small part of the nuclear wave packet reaches distances $R>3$ via evolution on the $1s\sigma_g$ curve. Thus the nuclear wave packet evolving along the $2p\sigma_{u}$ curve contributes much more to the low-energy KER peaks. As a result, only trajectories where the second jump takes place from the $2p\sigma_{u}$ state are shown in Fig. \ref{fig:kertot_1ev_6400nm}. A clear oscillatory dependence of the signal on the second jump time $T_{2}$ can be observed. This is because all the first jumps take place at the field extrema and there are enhancements when the second jumps occur near field extrema. The time differences between the second and first jumps are the evolution times within the H$_{2}^{+}$ system. The evolution times for the most prominent events (marked by red dots) for 21 and 15 laser cycles are about 6190 (150 fs) and 4420 (108 fs), which correspond to 7 and 5 laser cycles, respectively. In this manner, the MCWP approach allows us to find the typical time spent in the H$_{2}^{+}$ system, i.e., the time needed to ionize H$_{2}^{+}$, for each particular feature in the KER spectra. These times are long enough for the nuclear wave packets to reach quite large internuclear separations ($R>20$) along the dissociative $2p\sigma_{u}$ potential energy curve. As we will now show the difference (2 laser cycles for the two peaks at 1.2 eV and 1.5 eV in Fig. \ref{fig:kertot_wavelength_cycles}(d)) between the two evolution times allows us to obtain the dissociative kinetic energy $E_{d}$ for the low-energy peaks.\par 
 
One can see from Fig. \ref{fig:potential_curves} that there is nearly no variation in the $2p\sigma_{u}$ potential energy curve when $R>8$. Thus the dissociative kinetic energies of the low-energy protons are almost the same in spite of different numbers of laser cycles and ionization from different internuclear separations larger than 8. We can therefore estimate the dissociation energies in the following way. We consider two equations $E_{d}+{1}/{R_{21}}=1.2$ eV and $E_{d}+{1}/{R_{15}}=1.5$ eV, with $R_{21}$ and $R_{15}$ denoting the positions where the second ionization takes place when interacting with laser pulses with 21 and 15 laser cycles, respectively. In combination with the use of the relations $E_{d}=\frac{1}{2}mv^2$ and $vt=R_{21}-R_{15}$, we obtain the dissociative kinetic energies along the $2p\sigma_{u}$ potential energy curve from both the positions of the one- and three-photon resonances between the $2p\sigma_{u}$ and $1s\sigma_{g}$ curves. In these equations, $v$ is the velocity corresponding to $E_{d}$, $m$ here is the reduced mass of the nuclei, and $t$ is the evolution time difference for the 1.5 eV and 1.2 eV peaks. By substituting the evolution time difference of 2 laser cycles into the above equations, two physical solutions of the dissociative kinetic energy are obtained. One is 0.12 eV corresponding to the dissociative kinetic energy via the one-photon resonance. The other is 0.47 eV which corresponds to the dissociative kinetic energy after the three-photon resonance. Thus the potential energies at the one- and three-photon resonances can be expected as -13.6 eV + 0.12 eV = -13.48 eV and  -13.6 eV + 0.47 eV = -13.13 eV, where -13.6 eV is the energy of the dissociation limit. Both the energies obtained are quite close to the energies in the $2p\sigma_{u}$ curve of -13.46 eV and -13.24 eV at the one- and three-photon resonances as listed in Table \ref{tab:important_data}. In a similar way, we can obtain the potential energies of the $2p\sigma_{u}$ curve at the one- and three-photon resonances between the $2p\sigma_{u}$ and $1s\sigma_{g}$ curves for 3200 nm. 
\par
\begingroup
\squeezetable
\begin{table}[t!]
\begin{ruledtabular}
\centering
\caption{Data used for the analysis of the nuclear KER spectra following dissociative double ionization of H$_{2}$. The 750 nm case will be considered in Sec. \ref{sec:level4}. Here $\omega$ denotes the laser angular frequency. $R_{1}$ and $R_{3}$ are the positions of the one- and three-photon resonances between the $2p\sigma_{u}$ and $1s\sigma_{g}$ curves in H$_{2}^{+}$, respectively. $E_{u1}$ and $E_{u3}$ are the corresponding potential energies in the $2p\sigma_{u}$ potential energy curve at $R_{1}$ and $R_{3}$. $a_{0}$ is the Bohr radius. }
\begin{tabular} { c c c c c c c c }
 $\lambda$(nm) & $\hbar\omega$(eV) & $R_{1}(a_{0})$ & $R_{3}(a_{0})$ & $E_{u1}$(eV) & $E_{u3}$(eV) & ${1}/{R_{1}}$(eV) & ${1}/{R_{3}}$(eV)\\ [1ex] 
 \hline \\ [-1.5ex]
 800 & 1.55 & 4.74 & 3.28 & -12.69 & -10.7 & 5.7 & 8.3 \\
 1600 & 0.775 & 5.64 & 4.22 & -13.14 & -12.23 & 4.8 & 6.4 \\
 3200 & 0.388 & 6.5 & 5.12 & -13.35 & -12.91 & 4.2 & 5.3 \\
 6400 & 0.194 & 7.32 & 6 & -13.46 & -13.24 & 3.7 & 4.5 \\
 750 & 1.653 & 4.67 & 3.19 & -12.63 & -10.47 & 5.82 & 8.53\\
 \end{tabular}
 \label{tab:important_data}
\end{ruledtabular}
\end{table}
\endgroup

\begin{figure}[!ht]
  \includegraphics[width=0.35\textwidth]{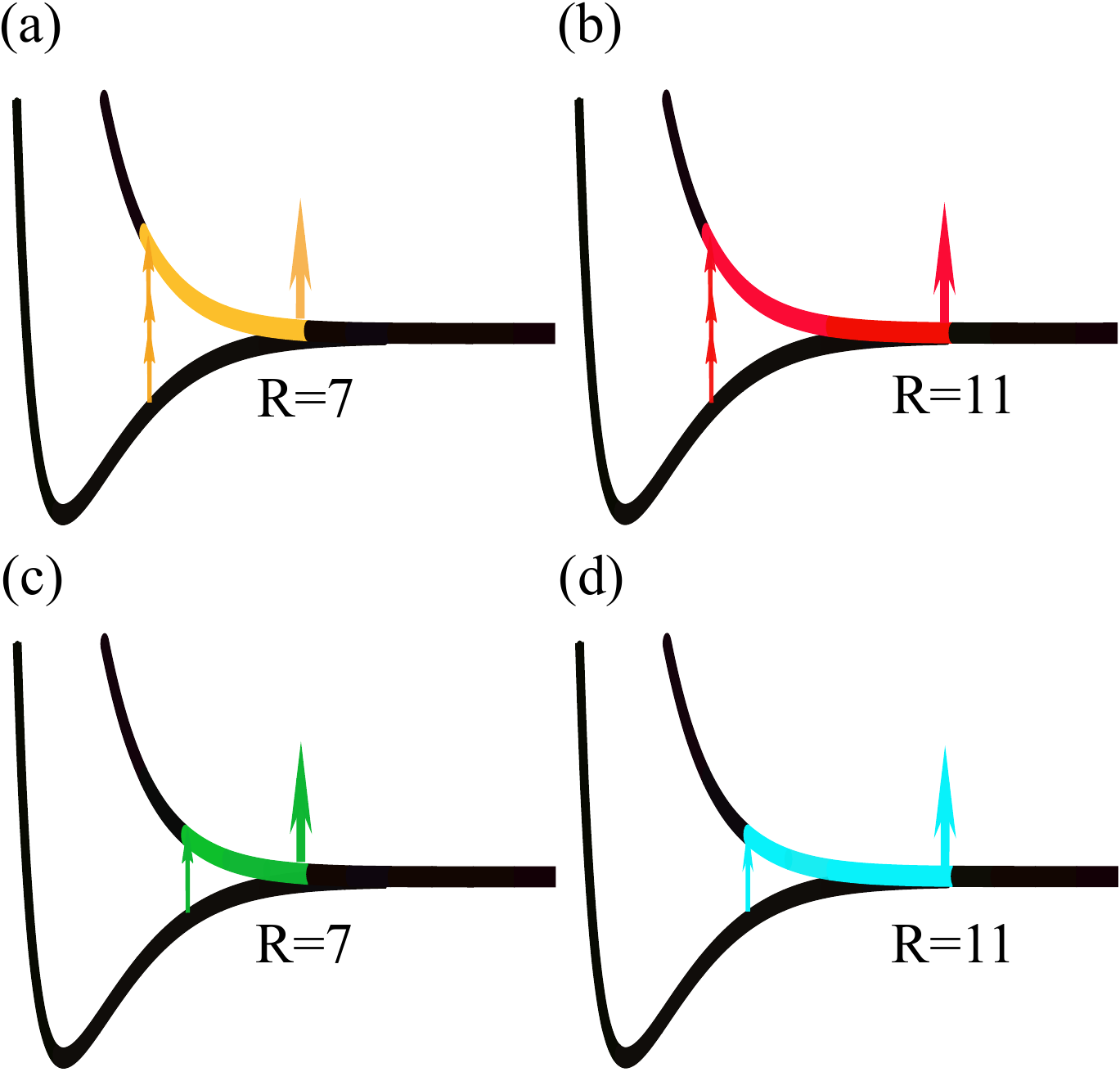}
  \centering
  \caption{Sketch of four processes of enhanced ionization at the two CREI positions of $R=7$ and $R=11$ (denoted by the outermost right arrows in the figures) after one- and three-photon resonance transitions. (a) and (b) CREI at $R=7$ and $R=11$ after the three-photon resonance. (c) and (d) CREI at $R=7$ and $R=11$ after the one-photon resonance. } 
  \label{fig:four_processes}
\end{figure}

A detailed analysis of the KER spectra in Fig. \ref{fig:kertot_wavelength_cycles} is conducted in the following with reference to Table \ref{tab:important_data}. The four processes shown in Fig. \ref{fig:four_processes} are expected to play an important role for the final nuclear KER spectra. They are representing enhanced ionization at the two CREI positions after the one- and three-photon resonant dipole coupling between the $1s\sigma_{g}$ and the $2p\sigma_{u}$ states in H$_{2}^{+}$ at smaller internuclear distances. In principle, ionization at the one- and three-photon resonances should also result in peaks close to their Coulomb repulsion energies as the dissociative kinetic energy is much smaller for these two cases. To better understand the origin of the peaks in Fig. \ref{fig:kertot_wavelength_cycles}, the positions of the one-photon resonance, $R_{1}$, and the three-photon resonance, $R_{3}$, are given in Table \ref{tab:important_data} for several wavelengths. In addition, the potential energies in the $2p\sigma_{u}$ potential energy curve ($E_{u1}$ and $E_{u3}$) and in the Coulomb potential energy curve for H$_{2}^{++}$ at $R_{1}$ and $R_3$ are given. \par

When the wavelength is 800 nm, as expected from Table \ref{tab:important_data}, there should be some enhancement in the proton yields at around 8.3 eV, 5.7 eV, 6.6 eV, 5.36 eV, 4.63 eV and 3.4 eV. They result from enhanced ionization at the three-photon resonance and the one-photon resonance as well as from the four processes pictured in Fig. \ref{fig:four_processes}. In Fig. \ref{fig:kertot_wavelength_cycles}(a),
the peaks around 4.7 eV mainly come from enhanced ionization at the CREI position of $R=7$ via the one-photon resonance, which can be verified by the following process: The dissociative kinetic energy is 13.43 eV - 12.69 eV = 0.74 eV, which is the difference in the $2p\sigma_{u}$ potential energy curve between $R_{3}=3.28$ and $R=7$. Adding this kinetic energy to the Coulomb repulsion energy ${1}/{7}$ (3.89 eV) at $R=7$, the total kinetic energy of the proton is 4.63 eV. Similarly, the peaks around 5.3 eV in Fig. \ref{fig:kertot_wavelength_cycles}(a) can be related to enhanced ionization at the CREI position of $R=11$ via the three-photon resonance, which are quite close to the expected 2.89 eV + 2.47 eV = 5.36 eV. The peaks around 7.3 eV are indeed from the combined contributions from at least two processes. These are enhanced ionization at the three-photon resonance (8.3 eV) and at the CREI position of $R=7$ via the three-photon resonance (6.6 eV). These two relative broad peaks may give a peak at around (8.3 eV + 6.6 eV) $/$ 2 = 7.45 eV. The peaks around 6.2 eV are mainly from the combination of enhanced ionization at the one-photon resonance (5.7 eV) and at the CREI position of $R=7$ via the three-photon resonance (6.6 eV). \par

When the wavelength is 1600 nm, Table \ref{tab:important_data} and reasoning as in the 800 nm case above shows that enhancements around 6.4 eV, 4.8 eV, 4.2 eV, 3.0 eV, 5.1 eV and 3.9 eV are expected to appear in the spectra shown in Fig. \ref{fig:kertot_wavelength_cycles}(b). As a result, the peaks around 6.4 eV mainly come from enhanced ionization via the three-photon resonance. The peaks around 5 eV are mainly from enhanced ionization via the one-photon resonance and enhanced ionization at $R=7$ after the three-photon resonance. The peaks around 4.3 eV are mainly from enhanced ionization at the CREI position of $R=7$ and enhanced ionization at $R=11$ via the three-photon resonance. Besides, large ionization at the CREI position of $R=11$ through the one-photon resonance leads to the peaks around 3 eV. \par

As the wavelength increases to 3200 nm, the expected enhancements in the spectra are around 5.3 eV, 4.2 eV, 4.0 eV, 2.8 eV, 4.4 eV and 3.2 eV, respectively [Table \ref{tab:important_data}]. The ionization at the three-photon resonance can result in peaks around 5.3 eV. The peaks around 4.0 eV mainly come from large ionization at the one-photon resonance or at the CREI position of $R=7$ via the one-photon resonance or even from enhanced ionization at $R=7$ after the three-photon resonance. The peaks around 2.6 eV come from large ionization at the CREI position of $R=11$ after the one-photon resonance, while the peaks around 3 eV are from enhanced ionization at $R=11$ after the three-photon resonance. Further increasing the wavelength to 6400 nm, there should be enhancement in the proton yields around five kinetic energies (4.5 eV, 3.7 eV, 2.6 eV, 4.1 eV and 2.9 eV [Table \ref{tab:important_data}]). There would be no events at $R=7$ after the one-photon resonance, as the position of the one-photon resonance $R=7.32$ is larger than $R=7$. The one-photon resonance and three-photon resonance give peaks around 3.6 eV and 4.9 eV. Enhanced ionization at the CREI position of $R=11$ from the one- and three-photon resonances give the respective peaks around 2.6 eV and 2.9 eV. 
 \par  
 
All of the plots in Figs. \ref{fig:kertot_wavelength_cycles}(a-d) are renormalized to better observe the common features in the spectra. There are large differences in the proton yields for the pulses. In Fig. \ref{fig:kertot_wavelength_cycles}(e), the total yields of protons as a function of the number of laser cycles under several wavelengths are shown in units of ionization probability per molecule, which are also the units for the nuclear KER spectra below. When the peak intensity of the laser pulse is fixed, the duration of the pulse does influence the total yields. The larger the duration, the larger the yield. 
Besides, for the same durations, e.g., 10 laser cycles at 1600 nm and 5 laser cycles at 3200, the yields are nearly identical. 

\section{\label{sec:level4} Delay dependence of KER spectra for H$_{2}$ interacting with two pulses}
\begin{figure}[!ht]
  \includegraphics[width=0.45\textwidth]{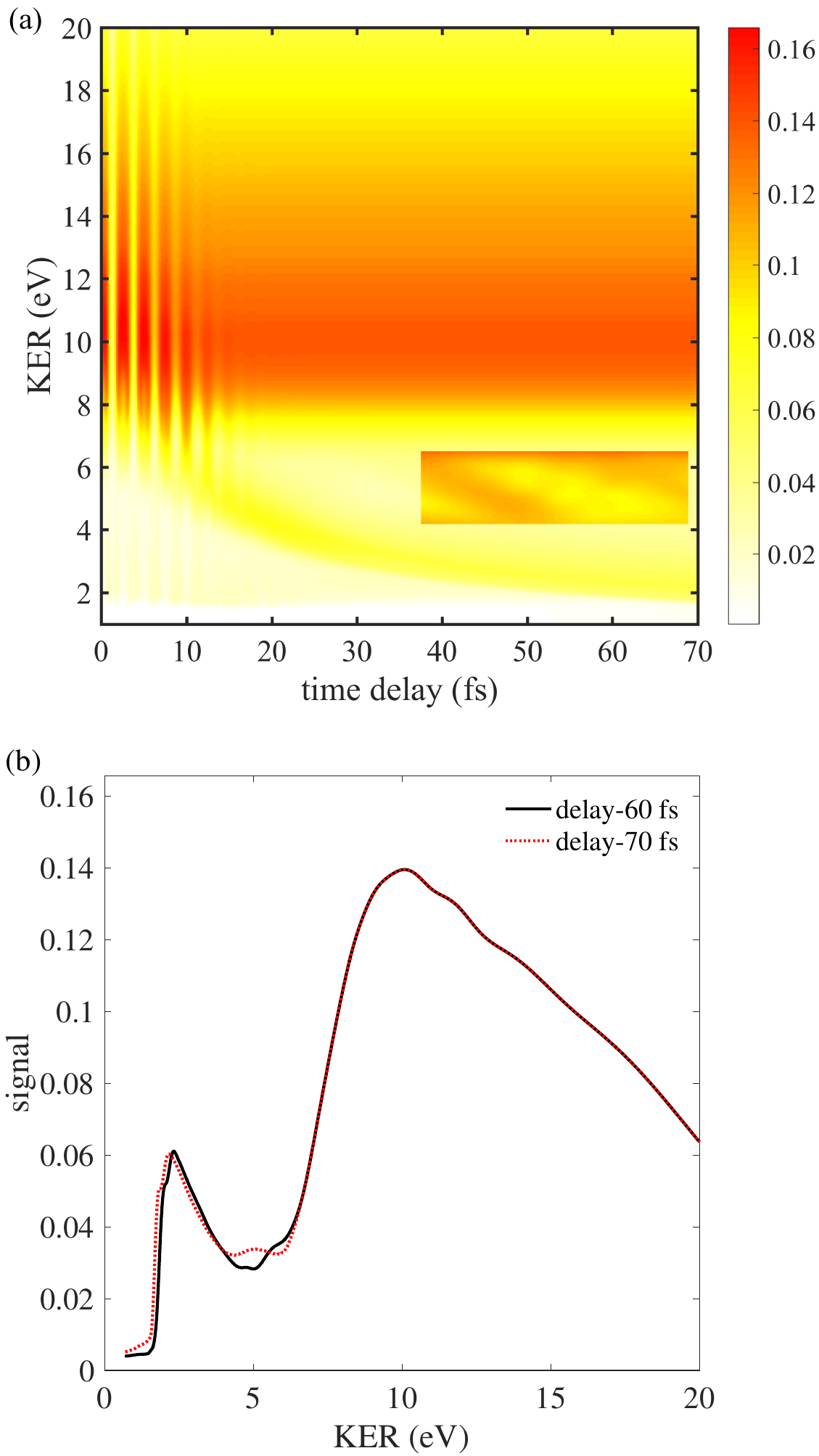}
  \centering
  \caption{(a) Nuclear KER distributions as a function of time delay between pump and probe pulses. The peak intensity of the pump (probe) pulse is $4\times 10^{14}$ $\text{W}/ \text{cm}^{2}$ ($6\times 10^{13}$ $\text{W}/ \text{cm}^{2}$). The two pulses have the same wavelength (750 nm) and the same duration (3 laser cycles). The data are plotted to the power of 0.2 in order to gain a larger visibility of the structures and after that the signal in the insert was multiplied by a factor of 3. (b) Cuts from (a) before multiplying by a factor 3 at time delays of 60 fs and 70 fs. } 
  \label{fig:pump_probe_750_3}
\end{figure}

 When molecules interact with two or more laser pulses, dynamics of the nuclei as well as dynamics of the photoelectrons can be studied as a function of the time delay between the pulses. Usually, the first pulse is the pump which stimulates electronic excitation and nuclear motion. The following pulse acts as the probe and is used to detect the nuclear or electronic wave packets. The pump-probe technique has been widely used to study ultrafast dynamics of both electrons and nuclei and we refer the readers to Refs. \cite{PhysRevLett.98.073003,PhysRevLett.97.103004,thumm2008time,bocharova2008direct, PhysRevLett.93.183202,xu2015experimental} for recent examples in H$_{2}$. In Fig. \ref{fig:pump_probe_750_3}(a), we plot the simulated KER distributions as a function of time delay between pump and probe laser pulses with their polarization axes parallel to the molecular axis. In order to test the performance of our approach, the parameters of the two laser pulses are chosen according to Ref. \cite{xu2015experimental}: The central wavelength and the FWHM duration of the two pulses are 750 nm and 3 laser cycles, respectively. The pump laser has a peak intensity of $4\times 10^{14}$ $\text{W}/ \text{cm}^{2}$ while the probe has a peak intensity of $6\times 10^{13}$ $\text{W}/ \text{cm}^{2}$. We find a good agreement between Fig. \ref{fig:pump_probe_750_3}(a) and the experimental results of Ref. \cite{xu2015experimental}. In particular, the high-energy branch around 10 eV and the energy-decreasing branch from 6 eV to 2 eV for increasing time delay are reproduced. \par
 
In Fig. \ref{fig:pump_probe_750_3}(a), the channel with KER around 10 eV mainly comes from double ionization of H$_{2}$ by the more intense pump pulse. This channel shows no delay dependence when the delay is larger than 20 fs. However, for delays smaller than 20 fs, oscillation of the KER distributions with a period of one laser cycle is observed. This oscillation is a signature of constructive and destructive interference between the two pulses when the two pulses overlap in time. The channel with KER in the 2-6 eV range has a strong delay dependence and is due to the ionization of H$_{2}^{+}$ by the relatively weak probe pulse. The main energy-decreasing branch from 6 eV to 2 eV for increasing delay is a result of the nuclear wave packet moving along the $2p\sigma_{u}$ potential energy curve in H$_{2}^{+}$. In addition, there are two much weaker energy-decreasing branches. A zoom of part of the branches is shown in the insert to the right of Fig. \ref{fig:pump_probe_750_3}(a). These branches can be shown to result from the oscillatory movement of the nuclear wave packet along the $1s\sigma_{g}$ potential energy curve in H$_{2}^{+}$: The duration between the first dominant decreasing branch starting at a time delay around 15 fs and the second branch highlighted in the insert is around 20-25 fs corresponding to a full oscillatory motion in the $1s\sigma_{g}$ curve. In order to extract properties of the $2p\sigma_{u}$ potential energy curve, we resort to solving the equations of the dissociative kinetic energy as discussed in Sec. \ref{sec:level3}. We use the peak positions of 2.17 eV and 2.34 eV for the two low-energy peaks at time delays of 60 fs and 70 fs [Fig. \ref{fig:pump_probe_750_3}(b)]. With this difference in evolution time of 10 fs, we obtain the dissociative kinetic energy of about 0.93 eV. Thus the $2p\sigma_{u}$ potential energy at the one-photon resonance position for 750 nm is about -13.6 eV + 0.93 eV = -12.67 eV, which is very close to the expectation of -12.63 eV from Table. \ref{tab:important_data}. \par

\begin{figure}[ht]
  \includegraphics[width=0.45\textwidth]{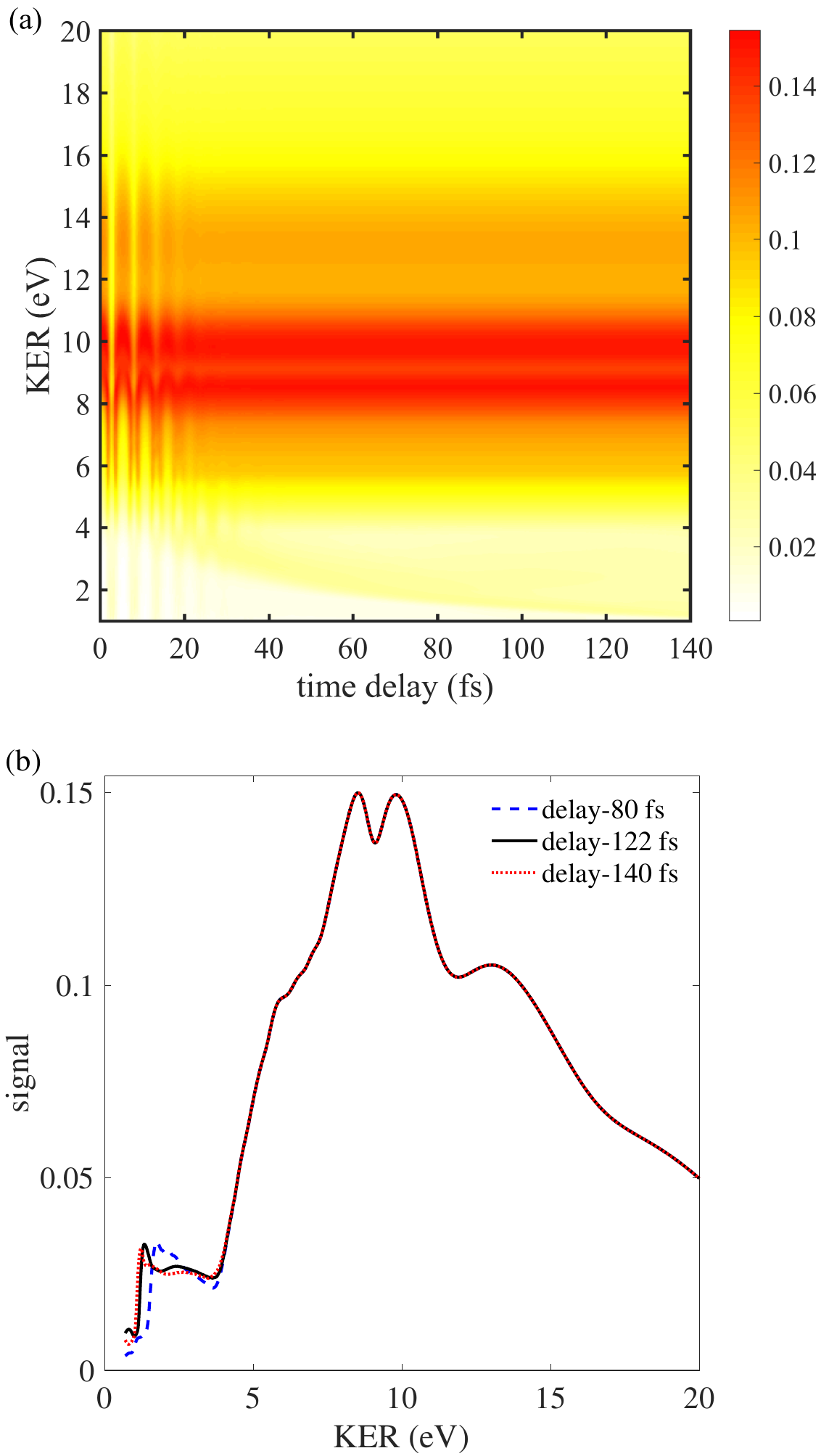}
  \centering
  \caption{(a) Nuclear KER distributions as a function of time delay between pump and probe pulses. The peak intensity of the pump (probe) pulse is $4\times 10^{14}$ $\text{W}/ \text{cm}^{2}$ ($6\times 10^{13}$ $\text{W}/ \text{cm}^{2}$). Both the wavelengths are 1600 nm and the duration of both pulses is 3 laser cycles. The data are plotted to the power of 0.2 in order to gain a larger visibility of the structures. (b) Cuts from (a) at time delays of 80 fs, 122 fs and 140 fs. } 
  \label{fig:pump_probe_1600_3}
\end{figure} 

As the wavelength of the laser pulses becomes larger, more structures in the KER spectra are expected as the larger laser duration can activate more ionization pathways. In Fig. \ref{fig:pump_probe_1600_3}(a), we plot the KER distributions as a function as delay, when the wavelengths of both pulses are 1600 nm while the other parameters are as in Fig. \ref{fig:pump_probe_750_3}. Processes similar to the ones in Fig. \ref{fig:pump_probe_750_3}(a) can be seen in Fig. \ref{fig:pump_probe_1600_3}(a). For example, interference between the two pulses at time delays shorter than 30 fs and movement of the nuclear wave packet along the $2p\sigma_{u}$ potential energy curve (branch decreasing from 5 eV). Three peaks around 10 eV are observed. The origin of these peaks is discussed below. As shown in Fig. \ref{fig:pump_probe_1600_3}(b), when the delay is 140 fs, the low-energy peak of the KER spectra is around 1.2 eV, and when the delay is 80 fs, the low-energy peak of the KER spectra is around 1.75 eV. By solving similar equations as in Sec. \ref{sec:level3}, we obtain a dissociative kinetic energy of 0.42 eV, which is used to obtain the potential energy along the $2p\sigma_{u}$ curve at the one-photon resonance position for 1600 nm. The result of -13.6 eV + 0.42 eV = -13.18 eV is in very good agreement with the energy of -13.14 eV from Table. \ref{tab:important_data}.   \par 

\begin{figure}[ht]
  \includegraphics[width=0.45\textwidth]{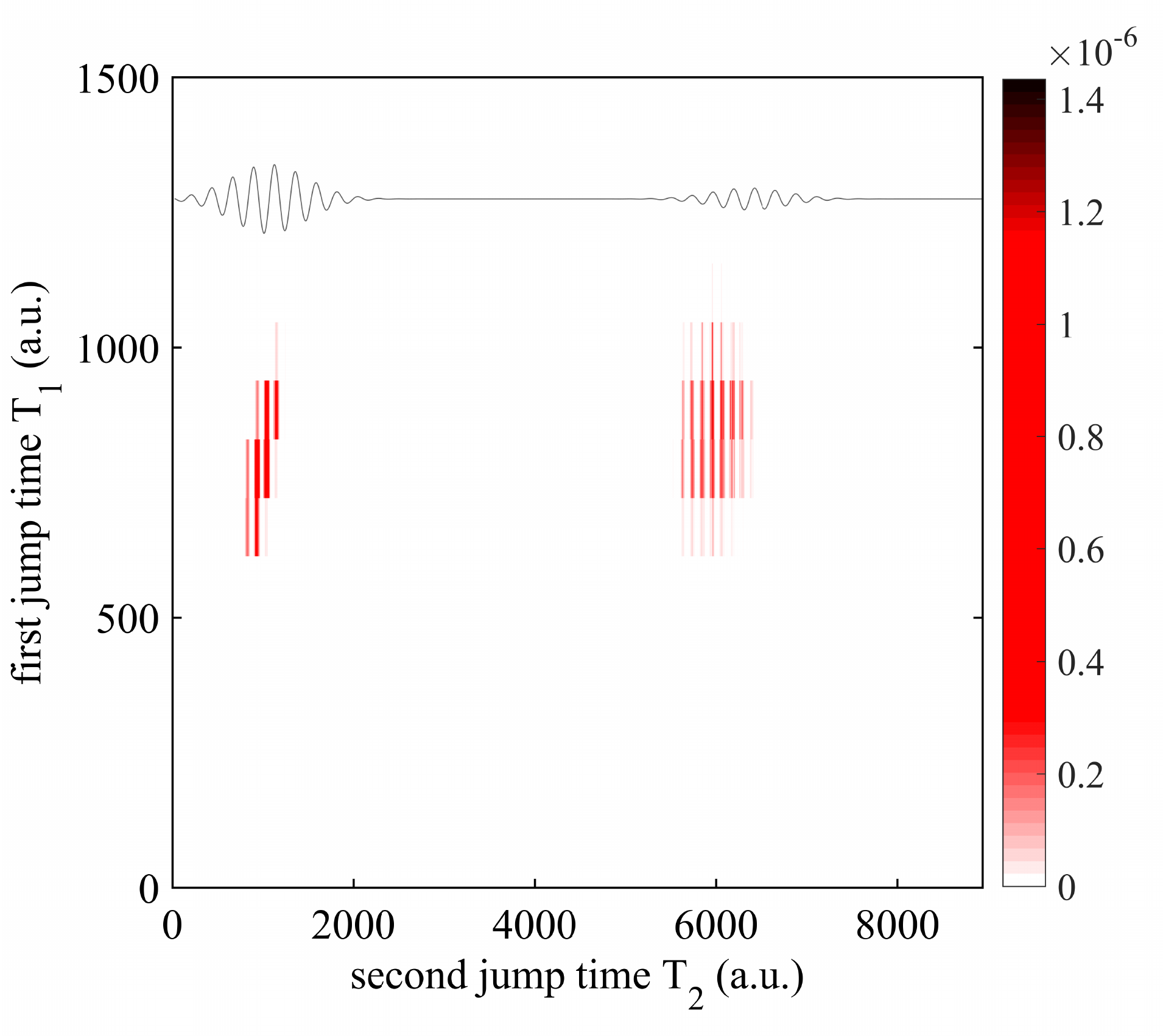}
  \centering
  \caption{ Contributions from different trajectories to the four peaks (8.5 eV, 9.8 eV, 13 eV and 1.4 eV) in the KER spectra in Fig. \ref{fig:pump_probe_1600_3}(b) when the delay is 122 fs. The distributions in the small $T_{2} \sim$ 1000 range result in three high-energy peaks around 8.5 eV, 9.8 eV and 13 eV while the distributions in the large $T_{2} \sim$ 6000 range give the low-energy peak around 1.4 eV. The two pulses are also shown by the grey line in the top of the figure. To show the distributions clearly in a single figure, signals for $T_{2} \sim 6000$ have been multiplied by a factor of 5000. } 
  \label{fig:ker_analysis_1600_3}
\end{figure} 

A trajectory analysis can help to clearly identify how the peaks in Fig. \ref{fig:pump_probe_1600_3}(b) are formed. In Fig. \ref{fig:ker_analysis_1600_3}, contributions from different trajectories to the four peaks (8.5 eV, 9.8 eV, 13 eV and 1.4 eV) in the KER spectra when the delay is 122 fs in Fig. \ref{fig:pump_probe_1600_3}(b) are shown. As in Fig. \ref{fig:kertot_1ev_6400nm}, a periodical dependence of the signal with the second jump time $T_{2}$ is clearly seen. The three peaks around 8.5 eV, 9.8 eV and 13 eV are mainly from double ionization induced by the pump pulse alone: The signals around 13 eV mainly come from trajectories in which the evolution time in H$_{2}^{+}$ (difference between the first and second jump times) is about half a laser cycle; the signals around 9.8 eV are from trajectories where the evolution time is around one laser cycle; and the signals around 8.5 eV are from trajectories where the evolution time is about one and a half laser cycles. The low-energy peak around 1.4 eV is from trajectories where the first ionization is induced by the pump pulse and the second by the probe pulse. In this manner, as was the case in Sec. \ref{sec:level3}, the MCWP approach offers a very direct look into the timescale of the ionization dynamics associated with a given feature in the KER spectra. In Fig. \ref{fig:ker_analysis_1600_3}, we also notice that the first jumps mainly occur at the same time around 750 (18.3 fs) for the three high-energy peaks. The pump field is so strong that nearly all of the singly-ionized molecules would be further ionized in two laser cycles. \par

\begin{figure}[ht]
  \includegraphics[width=0.45\textwidth]{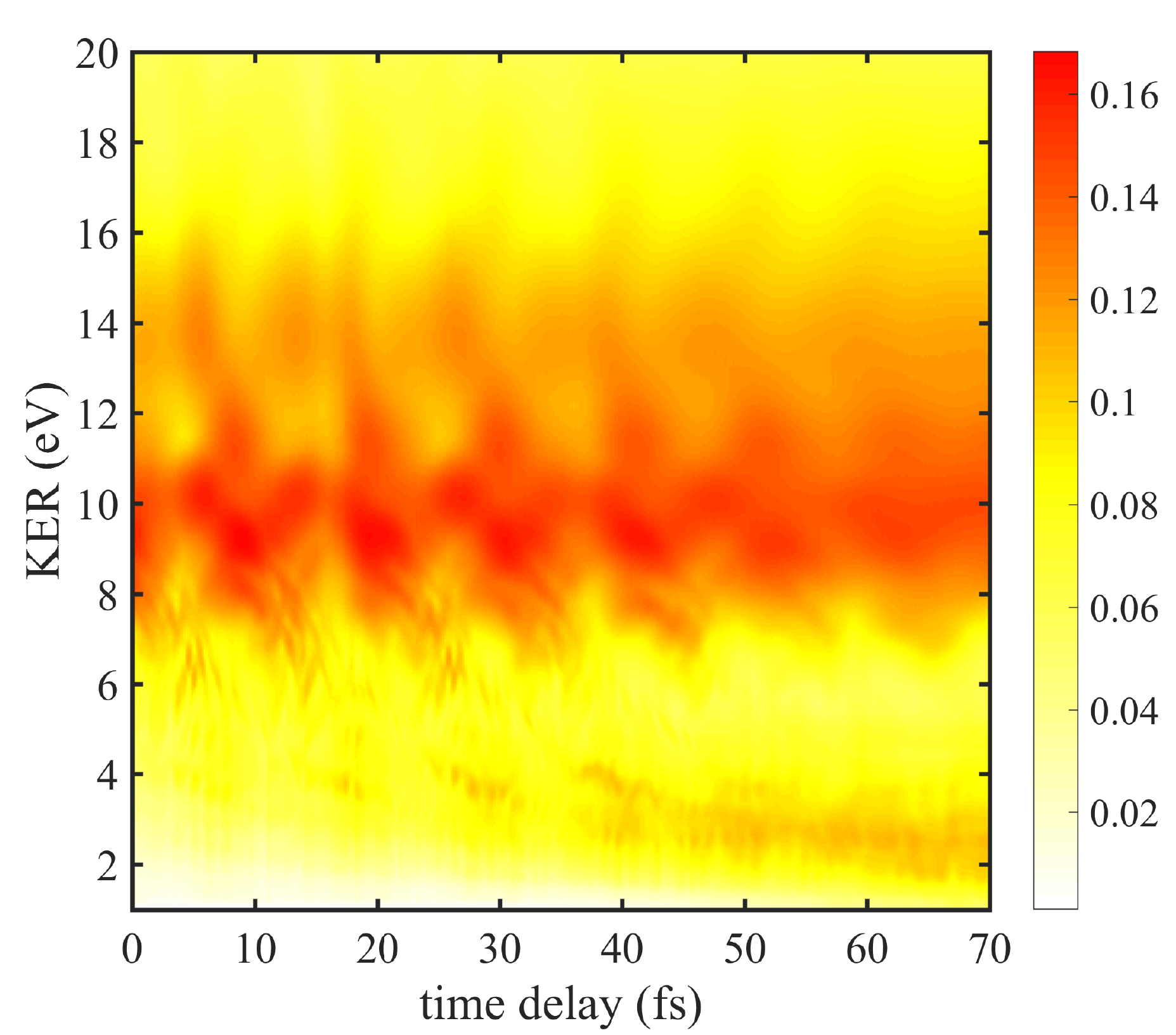}
  \centering
  \caption{ Nuclear KER distributions as a function of time delay between the pump and probe pulses. The peak intensity of the pump (probe) pulse is $4\times 10^{14}$ $\text{W}/ \text{cm}^{2}$ ($6\times 10^{13}$ $\text{W}/ \text{cm}^{2}$). The pump (probe) wavelength is 800 nm (6400 nm). The duration of both pulses is 3 laser cycles. The data are plotted to the power of 0.2 in order to gain a larger visibility of the structures.} 
  \label{fig:pump_probe_800_6400_3}
\end{figure} 

The evolution of the wave packet could be quite different when the wavelengths of the two pulses are different. In Fig. \ref{fig:pump_probe_800_6400_3}, the nuclear KER spectra of H$_{2}$ interacting with two pulses with different wavelengths are shown. The central wavelength of the pump pulse is 800 nm and that of the probe pulse is 6400 nm. They both contain three laser cycles. As the probe pulse is much longer, there are more structures and more yields in the KER spectra mainly induced by the probe pulse. Even when the delay reaches almost 70 fs, the overlap between the two pulses still plays a role in the final KER spectra, as is seen from the delay-dependence of the 10 eV channel. Moreover, there are more structures around the large kinetic energy channel at small delays as the effective duration of the pump pulse is becoming larger. Because of the complicated structures in the energy-decreasing branch (from 5 eV to 2 eV) in Fig. \ref{fig:pump_probe_800_6400_3}, the low-energy peaks in this case are not used to obtain the dissociative kinetic energy after the one-photon resonance. \par
\begin{figure}[ht]
  \includegraphics[width=0.38\textwidth]{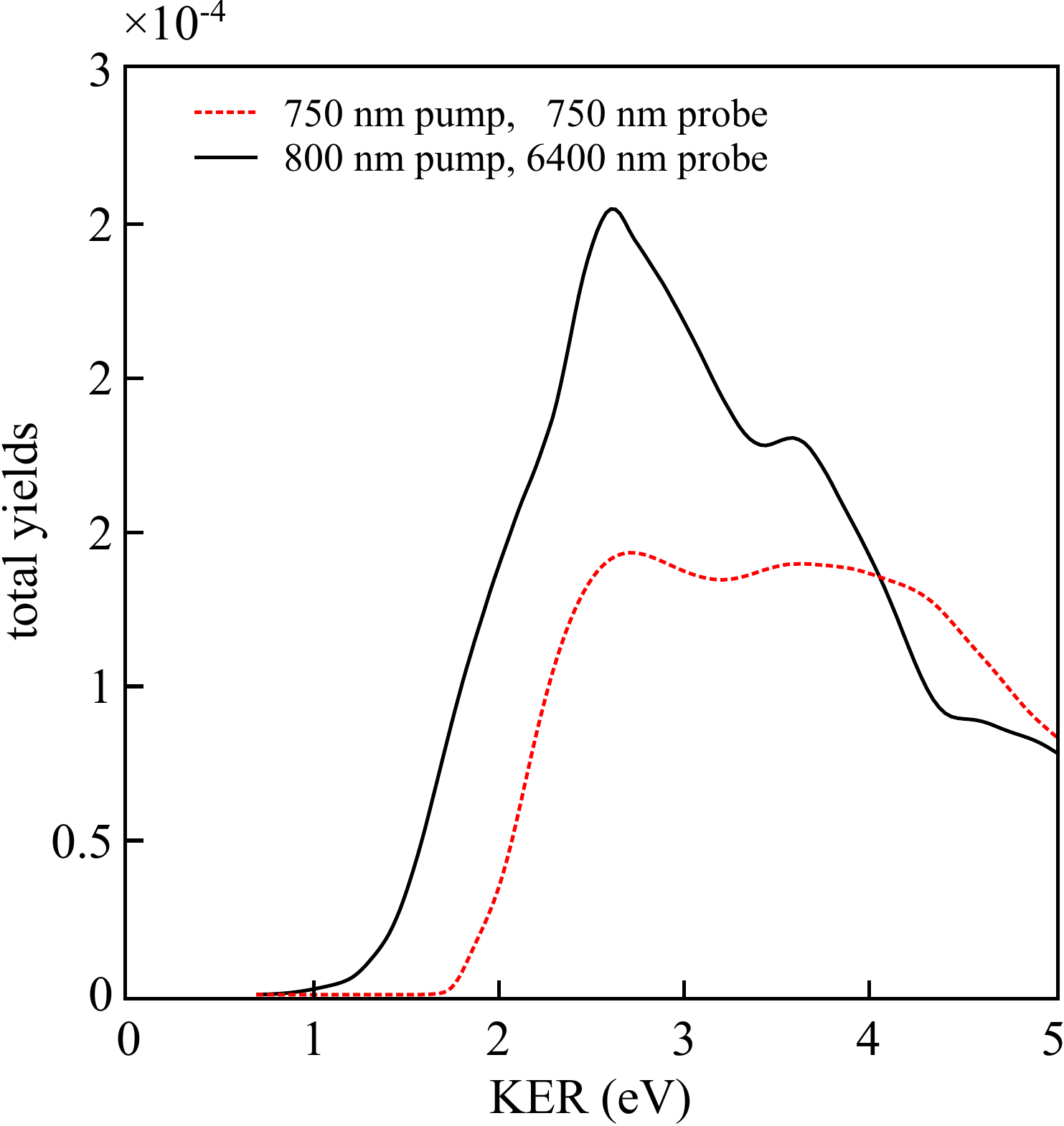}
  \centering
  \caption{ Total KER spectra integrated over all delays and shown as a function of KER energy. The red dashed line is for Fig. \ref{fig:pump_probe_750_3} and and the black solid line for Fig.  \ref{fig:pump_probe_800_6400_3}, respectively. } 
  \label{fig:total_counts_750_750_800_6400}
\end{figure} 

The pronounced energy-decreasing branch in Fig. \ref{fig:pump_probe_800_6400_3} is, however, useful for extracting information about CREI. The total KER spectra integrated over all delays in Figs. \ref{fig:pump_probe_750_3} and \ref{fig:pump_probe_800_6400_3} are shown in Fig. \ref{fig:total_counts_750_750_800_6400}, which shows a clear signature of the presence of the enhanced ionization. For both cases, the peaks around 3.6 eV mainly come from ionization at the CREI position of $R=11$ after the one-photon resonance between the $2p\sigma_{u}$ and $1s\sigma_{g}$ curves in H$_{2}^{+}$, while the energy-decreasing branch around 2.6 eV for delays larger than $\sim$ 40 fs represents nuclear KER spectra resulting from ionization from $R>11$. In theory, there should be peaks around 4.6 eV because of large ionization around the other CREI position at $R=7$ after the one-photon resonance between the $2p\sigma_{u}$ and $1s\sigma_{g}$ curves in H$_{2}^{+}$. However, they are not as clear as the peaks arond 3.6 eV. A possible origin is that enhanced ionization at around $R=7$ results in a broader energy peak than that at around $R=11$ because of the fact that there is a larger relative change in the Coulomb repulsion energy at smaller internuclear distances than at larger internuclear distances for a given $\Delta R$. \par 

The above discussion shows how the wavelength, pulse duration as well as delay of the laser pulses determines the nuclear KER spectra. The dynamics of the nuclear wave packets can be extracted when these parameters are well chosen. The ionization induced by the probe pulse after the one-photon resonance resulting from the pump pulse largely contributes to the energy-decreasing branch of the KER spectra. This delay-dependent channel will be more pronounced if the duration or the intensity of the probe pulse is increased. Meanwhile, when increasing the duration of the laser pulses, CREI would also show up more clearly on the averaged spectra integrated over all delays. 

\section{\label{sec:level5} Summary and Conclusion}
We calculated the nuclear KER spectra of H$_{2}$ interacting with laser pulses with central wavelengths ranging from the near-IR regime to the mid-IR regime. With increasing wavelength, we observe a shift in the nuclear kinetic energy distribution towards lower energies. A detailed analysis of the peaks in the KER spectra was carried out. Some of the peaks are from ionization taking place at the one- or three-photon resonance between the $1s\sigma_{g}$ and $2p\sigma_{u}$ curves in the H$_{2}^{+}$ system. Others are from CREI after the one- and three-photon resonances, or from a combination of these processes. Both for the cases of a single laser pulse and for two time-delayed pulses, characteristic low-energy peaks were observed when the duration or the delay of the pulses is large, e.g., for a single pulse with 21 laser cycles at 6400 nm and for two pump-probe pulses at 1600 nm with delay of 122 fs. The main physical origin of the low-energy peaks is ionization occurring at instants of field extrema when the nuclear wave packet has reached separations beyond the CREI position of $R=11$. For the single pulse case, the low-energy peak is a result of ionization after dissociation via the one- and three-photon resonances between the $1s\sigma_{g}$ and $2p\sigma_{u}$ states. For the two pulse case, it mainly originates from ionization following dissociation via the one-photon resonance between the two states. In the pump-probe simulation, by integrating the KER spectra with all delays, the effect of CREI was shown in the KER spectra. We also showed how to extract the dissociative energies of the $2p\sigma_{u}$ curve in H$_{2}^{+}$ following excitation at the one- or three- photon resonances at 750 nm, 1600 nm and 6400 nm. Finally, we illustrated in a few selected cases how the MCWP approach allows a trajectory analysis helping with the identification of dominant breakup pathways including an assessment of typical durations between the ionization events forming the characteristic features in the KER spectra.  
\par

\begin{acknowledgments}
We thank Lun Yue for help with numerical issues. This work was supported by the European Union Horizon 2020 research and innovation programme under the Marie Sklodowska-Curie grant agreement No. 641789 MEDEA (Molecular Electron Dynamics Investigated by Intense Fields and Attosecond Pulses), the European Research Council StG (Project No. 277767-TDMET), and the VKR center of excellence, QUSCOPE. The numerical results presented in this work were obtained at the Centre for Scientific Computing, Aarhus.
\end{acknowledgments}
%
\end{document}